\documentclass[aps,graphicx,6pt,showkeys,showpacs,onecolumn]{revtex4}
\usepackage{amsmath}
\usepackage{amscd}
\usepackage{graphicx}
\usepackage{subfigure}
\usepackage{appendix}

\begin{document}

\title[Short Title]{ Method for constructing shortcuts to adiabaticity by a substitute of counterdiabatic driving terms}
%\title[Short Title]{Flexible design of the adding Hamiltonian for shortcuts to adiabaticity}

\author{Ye-Hong Chen$^{1}$}
\author{Yan Xia$^{1,}$\footnote{E-mail: xia-208@163.com}}
\author{Qi-Cheng Wu$^{1}$}
\author{Bi-Hua Huang$^{1}$}
\author{Jie Song$^{2}$}

\affiliation{$^{1}$Department of Physics, Fuzhou University, Fuzhou 350002, China\\
             $^{2}$Department of Physics, Harbin Institute of Technology, Harbin 150001, China}

%\tableofcontents

\begin{abstract}
  We propose an efficient method to construct shortcuts to adiabaticity (STA) through designing a substitute Hamiltonian
  to try to avoid the defect that the speed-up protocols' Hamiltonian
  may involve the terms which are difficult to be realized in practice.
  We show that as long as the counterdiabitic coupling terms, even only some of them, have been nullified by the adding Hamiltonian,
  the corresponding shortcuts to adiabatic process could be constructed and the adiabatic process would be speeded up.
  As an application example, we apply this method to the popular Landau-Zener model for the realization of fast population inversion.
  The results show that in both Hermitian and non-Hermitian systems, we can design different adding Hamiltonians to replace the
  traditional counterdiabitic driving Hamiltonian to speed up the process.
  This method provides lots of choices to design the adding terms of the Hamiltonian such that
  one can choose the realizable model in practice.
\end{abstract}


\maketitle

Since Demirplack and Rice \cite{MDSARJpcJpc0308} and Berry
\cite{MVBJpamt09} proposed that the addition of a suitable
``counterdiabatic (CD)'' term $H_{cd}$ to an original time-dependent
Hamiltonian $H_{0}(t)$ can suppress transitions between
different time-dependent instantaneous eigenbasis of $H_{0}(t)$, an
emergent field named ``Shortcuts to adiabaticity'' (STA)
\cite{XCILARDGOJGMPra10,ETSISMGMMACDGOARXCJGMAmop13} which aims at
designing nonadiabatic protocols to speed up quantum adiabatic
process has been taken into our eyes and has attracted much interest
\cite{XCJGMPra12,AdCPrl13,Pra062116,ETSISMGMMACDGOARXCJGMAmop13,Pra13013415,Jpb43085509,Pra85033605,Pra84043434,Pra89012326,Lp24105201,CYH}.
To find shortcuts to adiabatic dynamics, several formal solutions
which are in fact strongly related or even potentially equivalent to
each other have been proposed, for instance, ``Counterdiabatic
driving'' \cite{XCILARDGOJGMPra10,AdCPrl13,Pra062116} (it can also
be named as ``Transitionless quantum driving'') and invariant-based
inverse engineering \cite{Pra062116,XCJGMPra12}. After years of
development, the theory of shortcuts to adiabatic dynamics gradually
becomes consummate, and STA has been applied in a wide range of
fields including ``fast cold-atom'', ``fast ion transport'', ``fast
expansions'', ``fast wave-packet splitting'', ``fast quantum
information processing'', and so on
\cite{XCJGMPra12,ETSISMGMMACDGOARXCJGMAmop13,Pra13013415,Jpb43085509,Pra85033605,Pra84043434,Pra89012326,Lp24105201,CYH,Jpb42241001,Prl104063002,Pra82053403,Njp13113017,
Pra043415,Njp14013031,Pra84031606Epl9660005,Njp14093040,Pra82033430,Epl9323001}.

Nevertheless, a problem has been always haunting in accelerating
adiabatic protocols: the structure or the values of the
shortcut-driving Hamiltonian might not exist in practice.
It is known to all that if the Hamiltonian is hard or even impossible to
be realized in practice, the protocols will be useless. In
view of that, %we are led to ask if it is possible to avoid this defect.
%That is also why in resent years, the several specific methods have been proposed to skillfully solve the problem.
%To overcome the problem, several specific methods have been proposed.
%
%In this scenario,
several ingenious methods that aim at amending the
problematic terms of the shortcut-driving Hamiltonian to satisfy the
experimental requirements have been proposed in recent years
\cite{Prl109100403,Pra89053408,Njp16015025,Pra90060301,Pra08743402,Pra89043408,Pra8705250289063412}.
For example, Ib\'{a}\~{n}ez \emph{et al.} \cite{Pra08743402}
examined the limitations and capabilities of superadiabatic
iterations to produce a sequence of STA in 2013. They calculated the
adding term by iteration method until the adding term was realizable
in practice, hence the problem could be avoided. Later, in 2014,
Mart\'{i}nez-Garaot \emph{et al.} \cite{Pra89043408} used the
dynamical symmetry of the Hamiltonian to find, by means of Lie
transforms, alternative Hamiltonians that achieved the same goals as
speed-up protocols did, while without directly using the CD Hamiltonian.
These ideas \cite{Pra08743402,Pra89043408,Pra8705250289063412}
inspire us that finding some substitute Hamiltonians for the
shortcut-driving Hamiltonian could be an efficient way to overcome
the problem that the speed-up protocols' Hamiltonian may involve the
terms which are difficult to be realized in practice.
Therefore, in this paper, by using reverse thinking, we come up
an idea to design an adding Hamiltonian which can also nullify the
nonadiabatic coupling term to achieve the same goals as the
shortcut-driving Hamiltonian does. Different from the previous works that
the adding term is calculated from the original Hamiltonian, we aim
at finding different ways to nullify the nonadiabatic coupling and
ensuring the shortcut-driving Hamiltonian can be realized in
practice.

The starting point is a time-dependent Hamiltonian $H_{0}(t)$ with
$N$ eigenstates $\{|\phi_{n}(t)\rangle\}$
\begin{eqnarray}\label{eq0}
  H_{0}(t)|\phi_{n}(t)\rangle=E_{n}(t)|\phi_{n}(t)\rangle.
\end{eqnarray}
The instantaneous eigenstates satisfy
\begin{eqnarray}\label{eq3}
  \langle{\phi}_{n}(t)|\phi_{m}(t)\rangle=\delta_{nm},
\end{eqnarray}
and the closure relation
\begin{eqnarray}\label{eq4}
  \sum_{n}|{\phi}_{n}(t)\rangle\langle\phi_{n}(t)|=I.
\end{eqnarray}

The dynamics of a system governed by Hamiltonian $H_{0}(t)$ is described by the Schr\"{o}dinger equation
\begin{eqnarray}\label{eq5}
  i\hbar\partial_{t}|\psi(t)\rangle=H_{0}(t)|\psi(t)\rangle.
\end{eqnarray}
%Similar to Transitionless quantum driving, we assume that there exists a Hamiltonian that drives $H_{0}(t)$'s adiabatic basis $|\phi_{n}(t)\rangle$ exactly
In general, $|\psi(t)\rangle$ is a column vector, and we can express
it as $|\psi(t)\rangle=\sum_{n}{a_{n}(t)|\mu_{n}\rangle}=[a_{1}(t),a_{2}(t),\cdots,a_{n}(t)]^{t}$, where
the superscript $t$ denotes the transpose,
$\{a_{n}(t)\}$ are the probability amplitudes of all the bare
(diabatic) states of the system, and $\{|\mu_{n}\rangle\}$ are the basis vectors satisfying
\begin{eqnarray}\label{eq51}
  \sum_{n}|\mu_{n}\rangle\langle\mu_{n}|=1,  \ \
  \langle\mu_{m}|\mu_{n}\rangle=\delta_{mn},  \ \
  |\mu_{m}\rangle\langle\mu_{n}|=\sigma_{mn},
\end{eqnarray}
where $\sigma_{mn}$ is a matrix, in which the matrix element are all zero except the $m$th line and the $n$th column is 1.
To study adiabatic passage, we can
transform the system into another picture whose bare states are the
adiabatic basis (the instantaneous eigenstates of $H_{0}$) with the
rotation matrix $R(t)$ which will be introduced in the following. In this picture, the dynamics of the system
is also described by Schr\"{o}dinger equation
\begin{eqnarray}\label{eq6}
  i\hbar\partial_{t}|\psi^{e}(t)\rangle=H_{0}^{e}(t)|\psi^{e}(t)\rangle,
\end{eqnarray}
where the superscript $e$ denotes the system is in the ``eigen picture'', and $|\psi^{e}(t)\rangle=[c_{1}(t),c_{2}(t),\cdots,c_{n}(t)]^{t}$.

To transform the quantum system from the Schr\"{o}dinger picture to the ``eigen picture'', the transformation equation is expressed as
$|\psi^{e}(t)\rangle=R^{\dag}|\psi(t)\rangle$, or in form of matrix,
\begin{eqnarray}\label{eq7}
  \left(
  \begin{array}{c}
    c_{1} \\
    c_{2} \\
    \vdots
  \end{array}
  \right)
  =
  \left(
   \begin{array}{ccc}
     S_{11} & S_{12} & \cdots \\
     S_{21} & S_{22} & \cdots \\
     \vdots & \vdots & \vdots
   \end{array}
  \right)
  \left(
  \begin{array}{c}
    a_{1} \\
    a_{2} \\
    \vdots
  \end{array}
  \right),
\end{eqnarray}
where $S_{mn}=\langle\phi_{m}|\mu_{n}\rangle$ and
\begin{eqnarray}\label{eq8}
  R^{\dag}(t)=
  \left(
   \begin{array}{ccc}
     S_{11} & S_{12} & \cdots \\
     S_{21} & S_{22} & \cdots \\
     \vdots & \vdots & \vdots
   \end{array}
  \right).
\end{eqnarray}
And we can also express the rotation matrix $R^{\dag}(t)$ as
\begin{eqnarray}\label{eq91}
  R^{\dag}=\sum_{m,n}\sigma_{mn}\langle\phi_{m}|\mu_{n}\rangle=\sum_{m,n}|\mu_{m}\rangle\langle\mu_{n}|\langle\phi_{m}|\mu_{n}\rangle=\sum_{m}|\mu_{m}\rangle\langle\phi_{m}|.
\end{eqnarray}
Putting this relationship into
eq. (\ref{eq5}) and eq. (\ref{eq6}), we obtain
\begin{eqnarray}\label{eq9}
  H_{0}^{e}(t)={R}^{\dag}H_{0}R-i\hbar {R}^{\dag}\dot{R},
\end{eqnarray}
where the dot means time derivative and
\begin{eqnarray}\label{eq10}
  {R}^{\dag}H_{0}R= \sum_{n}{\sigma_{nn} E_{n}},
\end{eqnarray}
is the diagonalization matrix for Hamiltonian $H_{0}(t)$, and
\begin{eqnarray}\label{eq11}
  i\hbar {R}^{\dag}\dot{R}&=&i\hbar\sum_{n}\sigma_{nn}\langle{\phi}_{n}(t)|\dot{\phi}_{n}(t)\rangle \cr\cr
                                 &&+i\hbar\sum_{n\neq m}\sigma_{nm}\langle{\phi}_{n}(t)|\dot{\phi}_{m}(t)\rangle.
\end{eqnarray}
As we can find, the integral of the first term in eq. (\ref{eq11}) is just the adiabatic phase, and the second term is the nonadiabatic coupling.
If $|\hbar\langle{\phi}_{n}(t)|\dot{\phi}_{m}(t)\rangle|\ll |E_{n}-E_{m}|$, then the transitions in the instantaneous eigenbasis
are suppressed and the evolution is adiabatic. That is what is called the adiabatic condition which limits the speed.
To construct shortcuts to speed up the dynamics, the convenient way is adding a Hamiltonian $H_{1}^{e}=i\hbar {R}^{\dag}\dot{R}$ to counteract
the nonadiabatic coupling. Moving back to the
Schr\"{o}dinger picture,
\begin{eqnarray}\label{eq12}
  H_{1}=RH_{1}^{e}{R}^{\dag}=i\hbar\dot{R}{R}^{\dag}=i\hbar \sum_{n}{|\dot{\phi}_{n}(t)\rangle\langle{\phi}_{n}(t)|}.
\end{eqnarray}
That is, we calculate the CD term through a different way from Berry's transitionless tracking algorithm.
In general, shortcuts can be constructed just by directly adding CD term in the original Hamiltonian $H_{0}(t)$.
However, as we mentioned above, such CD term always makes troubles in practice.
In this paper, we try to use reverse thinking to find other ways to nullify the nonadiabatic coupling.
In order to obtain a general result, we further assume that the instantaneous eigenstate
$|\phi_{n}(t)\rangle=[\phi_{n1},\phi_{n2},\phi_{n3},\cdots]^{t}$, where the time-dependent $\phi_{nm}$ denotes the $m$th
element of the column vector $|\phi_{n}(t)\rangle$.
Then, we assume that there exists a Hamiltonian $H_{add}=\sum_{k,l}\sigma_{kl}A_{kl}$.
It should be noted that to make sure adding Hamiltonian is practicable in practice,
it is better to choose the coefficients $A_{kl}$ to satisfy the condition $A_{nm}^{*}=A_{mn}$ ($n\neq m$) \cite{XCILARDGOJGMPra10,XCJGMPra12,Pra84023415,Pra8705250289063412,Pra08743402,Pra89053408,Njp14093040}.
By adding this Hamiltonian into eq. (\ref{eq9}), we obtain
\begin{eqnarray}\label{eq13}
  H^{e}=H_{0}^{e}+{R}^{\dag}H_{add}R,
\end{eqnarray}
in which
\begin{eqnarray}\label{eq13b}
  R^{\dag}H_{add}R=\sum_{n,m,k,l}\sigma_{nm}\phi^{*}_{nk}\phi_{ml}A_{kl}.
\end{eqnarray}
The term
${R}^{\dag}H_{add}R$ does not necessarily equal to $i\hbar
{R}^{\dag}\dot{R}$. So long as $R^{\dag}H_{add}R$ can nullify the
nonadiabatic coupling term $i\hbar\sum_{n\neq
m}\sigma_{nm}\langle{\phi}_{n}(t)|\dot{\phi}_{m}(t)\rangle$,
the shortcuts would be constructed. In other words, the shortcuts
will be constructed as long as
$\sum_{k,l}\phi^{*}_{nk}\phi_{ml}A_{kl}=i\hbar\langle\phi_{n}|\dot{\phi}_{m}\rangle$
($n\neq m$). In fact, the shortcuts are still constructible even
when only some of the terms in the matrix $i\hbar\sum_{n\neq
m}\sigma_{nm}\langle{\phi}_{n}(t)|\dot{\phi}_{m}(t)\rangle$
can be nullified. For example, if the terms
$\sigma_{n1}\langle\phi_{n}|\dot{\phi}_{1}\rangle$
are nullified, the transition $|\phi_{1}(t)\rangle\rightarrow
|\phi_{n\neq 1}(t)\rangle$ will be suppressed though the
transition $|\phi_{n\neq 1}(t)\rangle\rightarrow
|\phi_{1}(t)\rangle$ is allowed. In this way, the most important thing is
to make sure the initial state is perfectly in the eigenstate
$|\phi_{1}(t)\rangle$.

In the following, we take the two-level system as an example to
display the feasibility of the idea proposed above. We
assume a two-level Hermitian system has a ground level
$|1\rangle=[1,0]^{t}$ and an excited level
$|2\rangle=[0,1]^{t}$, its Hamiltonian in interaction picture is
given as
\begin{eqnarray}\label{eq14}
  H_{0}(t)=\frac{\hbar}{2}
                         \left(
                          \begin{array}{cc}
                            -\Delta(t) & \Omega(t)e^{-i\varphi(t)} \\
                            \Omega(t)e^{i\varphi(t)} & \Delta(t)
                          \end{array}
                         \right),
\end{eqnarray}
where $\Omega(t)$ is the Rabi frequency, assumed real, and $\Delta(t)$
is the detuning. The instantaneous eigenvectors for this system are $|\phi_{1}\rangle=\cos{\theta}e^{-i\varphi}|1\rangle-\sin{\theta}|2\rangle$
and $|\phi_{2}\rangle=\sin{\theta}|1\rangle+\cos{\theta}e^{i\varphi}|2\rangle$, where $\theta=\frac{1}{2}\arctan{\frac{\Omega}{\Delta}}$.
The corresponding eigenvalues are
$E_{1}=\frac{\hbar}{2}\sqrt{\Omega^{2}+\Delta^{2}}$ and $E_{2}=-\frac{\hbar}{2}\sqrt{\Omega^{2}+\Delta^{2}}$.
%Different adiabatic passage
%schemes correspond to $\Omega(t)$ and $\Delta(t)$ for which $|\psi(t)\rangle$ goes from $|1\rangle$ to $|2\rangle$ (or $|2\rangle$ to $|1\rangle$ ).
%The simplest example to implement this Hamiltonian is the Landau-Zener scheme with constant $\Omega(t)$
%and linear-in-time $\Delta(t)$. The second example is the Allen-Eberly (AE) scheme with
%$\Omega(t)=\Omega_{0}\text{sech}(\pi t/2t_{0})$ and $\Delta(t)=(2\beta^{2}t_{0}/\pi)\tanh{\pi/2t_{0}}$.
Then, the $R$ matrix can be given,
\begin{eqnarray}\label{eq15}
  R(\theta)=\left(
          \begin{array}{cc}
            \cos{\theta}e^{-i\varphi}  & \sin{\theta} \\
            -\sin{\theta} & \cos{\theta}e^{i\varphi}
          \end{array}
    \right),
 R^{\dag}(\theta)=
    \left(
          \begin{array}{cc}
            \cos{\theta}e^{i\varphi}  & -\sin{\theta} \\
            \sin{\theta} & \cos{\theta}e^{-i\varphi}
          \end{array}
    \right),
\end{eqnarray}
and
%the nonadiabatic coupling is
%\begin{small}
\begin{eqnarray}\label{eq16}
  i\hbar R^{\dag}\dot{R}=
     \hbar\left(
          \begin{array}{cc}
            \dot{\varphi}\cos^{2}{\theta}  & (i\dot{\theta}+\frac{\dot{\varphi}}{2}\sin{2\theta})e^{i\varphi} \\
            (-i\dot{\theta}+\frac{\dot{\varphi}}{2}\sin{2\theta})e^{-i\varphi} & -\dot{\varphi}\cos^{2}{\theta}
          \end{array}
    \right),
\end{eqnarray}
%\end{small}
where
\begin{eqnarray}\label{eq17}
  \dot{\theta}=\frac{\dot{\Omega}\Delta-\Omega\dot{\Delta}}{2(\Delta^{2}+\Omega^{2})}.
\end{eqnarray}
%On the other hand, we can easily obtain the only result of counterdiabatic term $H_{cd}$ by transitionless tracking algorithm (see ref. \cite{})
%\begin{eqnarray}\label{eq18}
%   H_{cd}=i\hbar\sum_{n}|\dot{\phi}_{n}\rangle\langle\phi_{n}|=\hbar
%          \left(
%            \begin{array}{cc}
%              \dot{\varphi}\cos^{2}{\theta}  & (i\dot{\theta}-\frac{\dot{\varphi}}{2}\sin{2\theta})e^{-i\varphi} \\
%              (-i\dot{\theta}-\frac{\dot{\varphi}}{2}\sin{2\theta})e^{i\varphi} & -\dot{\varphi}\cos^{2}{\theta}
%            \end{array}
%          \right).
%\end{eqnarray}
%It is without doubt that $H_{cd}=i\hbar \dot{R}{R}^{\dag}$.
%In previous work, the authors usually set $\varphi=0$ for simplicity,
%which means the geometric phase should be 0. However,  Is that possible to construct shortcuts without abandon the geometric phase?
%In view of that we want to find other adding Hamiltonians that play the same role with $H_{cd}$ but in different form.
According to transitionless tracking algorithm, the adding Hamiltonian (the CD Hamiltonian) is
%\begin{small}
\begin{eqnarray}\label{eq18}
   H_{cd}=i\hbar\sum_{n}|\dot{\phi}_{n}\rangle\langle\phi_{n}|
         =\hbar
          \left(
            \begin{array}{cc}
              \dot{\varphi}\cos^{2}{\theta}  & (i\dot{\theta}-\frac{\dot{\varphi}}{2}\sin{2\theta})e^{-i\varphi} \\
              (-i\dot{\theta}-\frac{\dot{\varphi}}{2}\sin{2\theta})e^{i\varphi} & -\dot{\varphi}\cos^{2}{\theta}
            \end{array}
          \right),
\end{eqnarray}
%\end{small}
which has been well known and might cause troubles in practice
(especially in multi-level and multi-qubit systems).
%If there exists
%another Hamiltonian which can also nullify the nonadiabatic coupling
%term, it might be a way to overcome the problem. So, the problem now becomes how to find another Hamiltonian
%which can play the same role as the CD Hamiltonian. That is what we will do in the following.
In order to tackle the problem, it might be wise to find another Hamiltonian which can also nullify
the nonadiabatic coupoling term and play the same role as the CD Hamiltonian.
We start from assuming an adding
Hamiltonian $H_{add}$ which is given with unknown parameters (we
have not made any hypothesis to the Hamiltonian here)
\begin{eqnarray}\label{eq19}
  H_{add}=
    \left(
          \begin{array}{cc}
            A_{11} & A_{12} \\
            A_{21} & A_{22}
          \end{array}
    \right).
\end{eqnarray}
It should notice that, though there are many choices for the coefficients $A_{mn}$, the premise should be
$H_{add}$ is realizable in practice. So when the coefficients are deduced, we should go back and check whether the adding
Hamiltonian is realizable or not. For example, in a two-level atomic system, it is better to set
$A_{12}$=$A_{21}^{*}$, and the boundary conditions (the phases are considered as 0 for convenience)
\begin{eqnarray}
  \text{Re}A_{12}=\text{const},\ \ \text{or} \ \ \text{Re}A_{12}(\tau)=\text{Re}A_{12}(t_{f})=0
\end{eqnarray}
and
\begin{eqnarray}
  \text{Im}A_{12}=\text{const}, \ \ \text{or} \ \
  \text{Im}A_{12}(\tau)=\text{Im}A_{12}(t_{f})=0,
\end{eqnarray}
where $\tau$ is the initial time and $t_{f}$ is the final time.

Then, according to eq. (\ref{eq13b}), we obtain
\begin{eqnarray}\label{eq20}
  R^{\dag}H_{add}R&=&
            \sigma_{11}[A_{11}\cos^{2}{\theta}+A_{22}\sin^{2}{\theta}-(A_{12}e^{i\varphi}+A_{21}e^{-i\varphi})\frac{\sin{2\theta}}{2}]      \cr\cr
            &&+\sigma_{12}[(A_{11}-A_{22})e^{i\varphi}\frac{\sin{2\theta}}{2}+A_{12}e^{2i\varphi}\cos^{2}{\theta}-A_{21}\sin^{2}{\theta}]   \cr\cr
            &&+\sigma_{21}[(A_{11}-A_{22})e^{-i\varphi}\frac{\sin{2\theta}}{2}-A_{12}\sin^{2}{\theta}+A_{21}e^{-2i\varphi}\cos^{2}{\theta}]  \cr\cr
            &&+\sigma_{22}[A_{11}\sin^{2}{\theta}+A_{22}\cos^{2}{\theta}+(A_{12}e^{i\varphi}+A_{21}e^{-i\varphi})\frac{\sin{2\theta}}{2}].
\end{eqnarray}
It is obvious that, as long as
\begin{eqnarray}\label{eq21}
  (A_{11}-A_{22})e^{i\varphi}\frac{\sin{2\theta}}{2}+A_{12}e^{2i\varphi}\cos^{2}{\theta}-A_{21}\sin^{2}{\theta}=\hbar(i\dot{\theta}+\frac{\dot{\varphi}}{2}\sin{2\theta})e^{i\varphi},
\end{eqnarray}
or
\begin{eqnarray}\label{eq22}
  (A_{11}-A_{22})e^{-i\varphi}\frac{\sin{2\theta}}{2}-A_{12}\sin^{2}{\theta}+A_{21}e^{-2i\varphi}\cos^{2}{\theta}=\hbar(-i\dot{\theta}+\frac{\dot{\varphi}}{2}\sin{2\theta})e^{-i\varphi},
\end{eqnarray}
the transition $|\phi_{2}(t)\rangle\rightarrow |\phi_{1}(t)\rangle$
or $|\phi_{1}(t)\rangle\rightarrow |\phi_{2}(t)\rangle$ is
suppressed, and the shortcut is constructed. These two eqs
(\ref{eq21}-\ref{eq22}) are the key points to realize the
accelerating adiabatic protocol. They determine the condition to be satisfied
to nullify the counterdiabitic coupling terms.
According to eqs. (\ref{eq21}-\ref{eq22}), we can pick out the
corresponding parameters to design $H_{add}$. A simple choice is to
set
\begin{eqnarray}\label{eq22a}
A_{11}&=&-A_{22}=\hbar\eta, \cr
A_{12}&=&A_{21}^{*}=\hbar(\alpha+i\beta)e^{-i\varphi},
\end{eqnarray}
where $\alpha,\beta,\eta$ are real, to ensure $H_{add}$ Hermitian.
Putting $\{A_{nm}\}$ into eqs. (\ref{eq21}-\ref{eq22}), we
obtain $\beta=\dot{\theta}$ and
$\alpha\cot(2\theta)+\eta=\dot{\varphi}/2$. Then, we have
\begin{eqnarray}\label{eq23}
    H^{e}&=&H_{0}^{e}+R^{\dag}H_{add}R \cr\cr
         &=&
         \hbar\left(
          \begin{array}{cc}
            E_{1}/\hbar+\chi(t) &
            0 \\
            0 &
            E_{2}/\hbar-\chi(t)
          \end{array}
         \right),
\end{eqnarray}
where $\chi(t)=\eta\cos{2\theta-\alpha\sin{2\theta}}-\dot{\varphi}\cos^{2}\theta$.
Hence, if the system's initial state is $|\psi(\tau)\rangle=[a_{1}(\tau),a_{2}(\tau)]^{t}$, then
\begin{eqnarray}\label{eq24}
  c_{1}(\tau)&=&a_{1}(\tau)\cos{\theta(\tau)}e^{i\varphi(\tau)}-a_{2}(\tau)\sin{\theta(\tau)}, \cr
  c_{2}(\tau)&=&a_{1}(\tau)\sin{\theta(\tau)}+a_{2}(\tau)\cos{\theta(\tau)}e^{-i\varphi(\tau)}.
\end{eqnarray}
By using the Schr\"{o}dinger equation (\ref{eq6}), we obtain
\begin{eqnarray}\label{eq25}
  i\hbar\partial_{t}
        \left(
           \begin{array}{c}
             c_{1}(t) \\
             c_{2}(t)
           \end{array}
        \right)
        =
        H^{e}
        \left(
           \begin{array}{c}
             c_{1}(t) \\
             c_{2}(t)
           \end{array}
        \right)
        \Rightarrow
        \left(
           \begin{array}{c}
             c_{1}(t)=c_{1}(\tau)e^{-i\int_{\tau}^{t}{E_{1}(t')/\hbar+\chi(t')dt'}} \\
             c_{2}(t)=c_{2}(\tau)e^{-i\int_{\tau}^{t}{E_{2}(t')/\hbar-\chi(t')dt'}}
           \end{array}
        \right).
\end{eqnarray}
That means the probability amplitudes $c_{1}(t)$ and $c_{2}(t)$ at time $t$ keep the same as that at the time $\tau$ with only phase difference.
Moving back to the interaction picture, the final state is
\begin{eqnarray}\label{eq26}
  |\psi(t)\rangle=
                   \left(
                      \begin{array}{c}
                        c_{1}(\tau)\cos{\theta}e^{-i\int_{\tau}^{t}{E_{1}(t')/\hbar+\chi(t')dt'}}e^{-i\varphi(t)}+c_{2}(\tau)\sin{\theta}e^{-i\int_{\tau}^{t}{E_{2}(t')/\hbar-\chi(t')dt'}}  \\
                        c_{2}(\tau)\cos{\theta}e^{-i\int_{\tau}^{t}{E_{2}(t')/\hbar-\chi(t')dt'}}e^{i\varphi(t)}-c_{1}(\tau)\sin{\theta}e^{-i\int_{\tau}^{t}{E_{1}(t')/\hbar+\chi(t')dt'}}
                      \end{array}
                   \right).
\end{eqnarray}
It is worth noting that when $\alpha=-\dot{\varphi}/2\sin{2\theta}$, $H_{add}=H_{cd}$.
In other words, the CD Hamiltonian calculated by transitionless tracking algorithm is one of the cases of the present method.

The idea can also be extended to the non-Hermitian systems. Assuming $H_{add}$ is a non-Hermitian Hamiltonian, for example,
the parameters are set as
\begin{eqnarray}\label{eq26a}
  A_{11}&=&-A_{22}=\hbar(\eta+i\gamma),   \cr
  A_{12}&=&A_{21}^{*}=\hbar(\alpha+i\beta)e^{-i\varphi},
\end{eqnarray}
where $\alpha,\ \beta,\ \eta$, and $\gamma$ are all real.
The choice of $A_{11}$ and $A_{22}$ here is just a relatively suitable example, we can also choose them as $\{A_{11}=\hbar(\eta+i\gamma),\ A_{22}=\hbar(\eta-i\gamma)\}$,
or $\{A_{11}=2\hbar(\eta+i\gamma),\ A_{22}=0\}$, or others as long as Im$(A_{11}-A_{22})\neq0$. Then, by solving the eq. (\ref{eq21}), we obtain
$\beta+\gamma\sin{2\theta}=\dot{\theta}$ and $\alpha\cot{2\theta}+\eta=\frac{\dot{\varphi}}{2}$.
While by solving eq. (\ref{eq22}), the result is quite different: $\beta-\gamma\sin{2\theta}=\dot{\theta}$
and $\alpha\cot{2\theta}+\eta=\frac{\dot{\varphi}}{2}$. That means if the adding Hamiltonian is non-Hermitian, we can not
ideally offset all the nondiagonal terms in eq. (\ref{eq16}). Only one of the two transition directions between the instantaneous eigenbasis
$|\phi_{1}\rangle$ and $|\phi_{2}\rangle$ can be forbidden.
That is to say, for the non-Hermitian system,
the initial state of the system should be ideally in one of the eigenstates, i.e., $|\psi(\tau)\rangle=\cos\theta(\tau) e^{-i\varphi(\tau)}|1\rangle-\sin\theta(\tau)|2\rangle$, hence,
$c_{1}=1$ and $c_{2}=0$. Then, the evolution of the system is described as
\begin{eqnarray}\label{eq27}
  &&i\hbar
  \left(\begin{array}{c}
          c_{1}(t) \\
          c_{2}(t)
        \end{array}
  \right)
  =
  H^{e}\left(\begin{array}{c}
          c_{1}(t) \\
          c_{2}(t)
        \end{array}
  \right)\cr\cr\cr&\Rightarrow&
  \left(\begin{array}{c}
          c_{1}(t)=\exp[-i\int_{\tau}^{t}{E_{1}/\hbar+(\eta+i\gamma)\cos{2\theta}-\alpha\sin{2\theta}-\dot{\varphi}\cos^{2}{\theta}}dt'] \\
          c_{2}(t)=0
        \end{array}
  \right).
\end{eqnarray}
We find that there is a real part in the exponential term which
may cause the decay. So, it would be better if we can make
$\int_{\tau}^{t}\gamma\cos{2\theta}=0$. A simple way is imposing
$\gamma\cos{2\theta}$ to be an odd function of time and assuming
$t_{f}=-\tau$ ($t_{f}$ is the total evolution time). The feature of
this method in the non-hermitian model is that the STA is sensitive
to the initial condition of the system. The initial state should be
ideally generated in the eigenstate which will not transfer to
others. It should be noticed here that, the imaginary part of diagonal terms
usually denotes the decay of the system. In most cases,
$\gamma$'s form is decided by the system so that we can not design it
as desired.
%Because of this, in ref. \cite{Pra8705250289063412}, the authors
%used optical waveguide system as the model to discuss Non-Hermitian STA.
However, this would not affect the feasibility of the present method, because in this paper, $\gamma$ would not be limited to some fixed form. It
can be any arbitrary function so long as the corresponding $\beta$
is realizable, for instance, $\gamma=\text{const}$, then
$\beta=\dot{\theta}\pm\gamma\sin{2\theta}$. This merit may be helpful in non-Hermitian systems which have been devoting an increasing interest
and have been discussed in recent years
\cite{Cjp561007,Jpa45444027}, for example,
the $\mathcal{PT}$-symmetric system \cite{Pra83052125,Pra86033813}.

\begin{figure}
 \scalebox{0.6}{\includegraphics {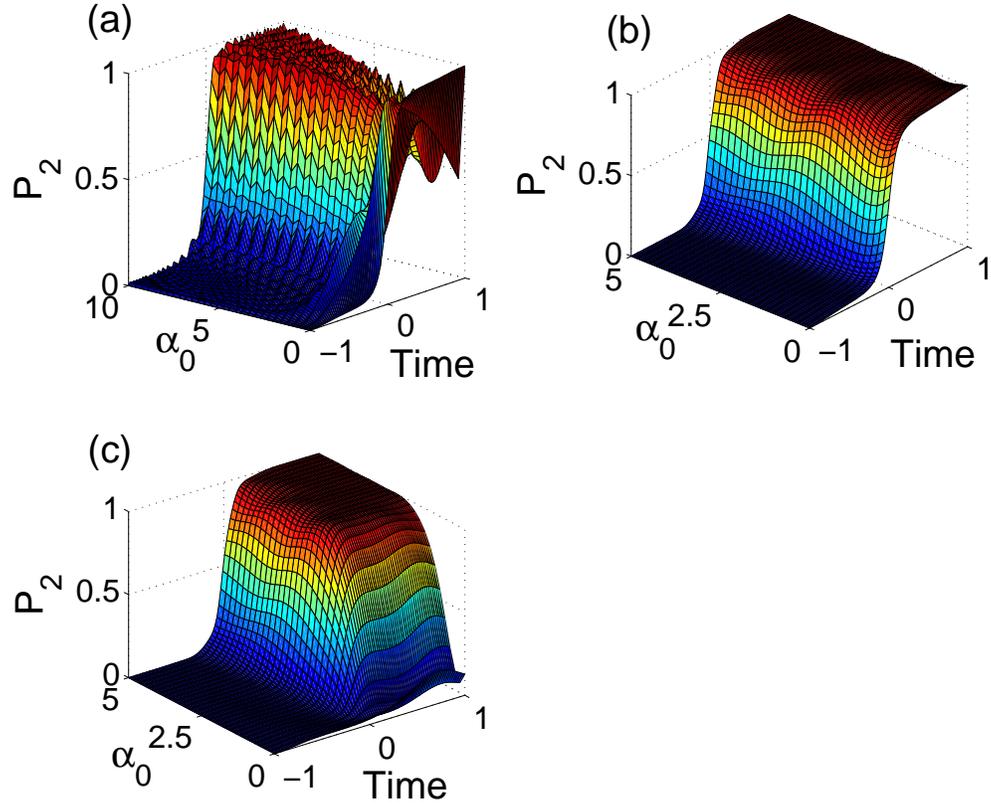}}
 \caption{
    The time-dependent $P_{2}$ versus $\alpha_{0}$ (in units $\Omega_{0}$) when $\{\varphi=0,\ \zeta=3\Omega_{0},\ t_{f}=1/\Omega_{0}\}$:
    (a) $\alpha=\alpha_{0}$ is const;
    (b) $\alpha=\alpha_{0}\dot{\theta}$ is time-dependent.
    (c) $\alpha=\alpha_{0}\dot{\theta}$ is time-dependent and $\beta=0$.
    The evolution time in the figure is in units of $1/\Omega_{0}$.}
 \label{fig1}
\end{figure}

Different adiabatic passage
schemes correspond to $\Omega(t)$ and $\Delta(t)$ for the system evolute from one bare state to the other. The simplest one is
the Landau-Zener scheme with constant $\Omega(t)$ and linear-in time $\Delta(t)$:
\begin{eqnarray}\label{eq28}
  \Omega(t)=\Omega_{0},\ \ \Delta(t)=\zeta^{2}t.
\end{eqnarray}
In this case, $\dot{\theta}=-\Omega_{0}\zeta^{2}/[2(\Omega_{0}^{2}+\zeta^{4}t^{2})]$.
The adding Hamiltonian $H_{add}$ is given as
\begin{eqnarray}\label{eq29}
  H_{add}=\hbar\left(
                 \begin{array}{cc}
                    \frac{\dot{\varphi}}{2}-\alpha\cot{2\theta}+i\gamma & (\alpha+i\dot{\theta}-i\gamma\sin{2\theta})e^{-i\varphi} \\
                   (\alpha-i\dot{\theta}+i\gamma\sin{2\theta})e^{i\varphi} & \alpha\cot{2\theta}-\frac{\dot{\varphi}}{2}-i\gamma
                 \end{array}
               \right).
\end{eqnarray}
Firstly, we discuss the situation when $\gamma=0$ (the system is Hermitian).
%There are two typical choices of $\alpha$ are $\alpha=0$ and $\alpha=-\frac{\dot{\varphi}}{2}\sin{2\theta}$.
%We would like to discuss how $\varphi$ affect the shortcuts at first,
In the interest of the effect of $\alpha$'s on STA, we set $\varphi=0$ in this part.
Two kinds of $\alpha$ will be discussed by numerical simulation.
(1): $\alpha$ is time-independent. Fig. \ref{fig1} (a) shows the
time-dependent population of the target state $|2\rangle$ ($P_{2}$)
versus $\alpha$ when the initial state is $|1\rangle$ and
$\{\varphi=0,\ \zeta=3\Omega_{0},\ t_{f}=1/\Omega_{0}$\}. The result
shows that in most of the cases, the shortcut could be
constructed successfully and the populations could
be transferred to the target state in a very short time. The
oscillation is caused by the diagonal term in eq. (\ref{eq29}).
%We know the geometric phase is a intriguing property of time-dependent quantum system. Therefore, analyzing the
%method when $\varphi\neq 0$
(2): $\alpha$ is time-dependent. For convenience, we choose
$\alpha=\alpha_{0}\dot{\theta}$ ($\alpha_{0}$ is time-independent).
As shown in Fig. \ref{fig1} (b), a nearly perfect population
transfer from $|1\rangle$ to $|2\rangle$ is realizable with
arbitrary $\alpha_{0}$. What is more, according to eq. (\ref{eq29}),
it is obvious when $\alpha_{0}$ is large enough,
$\alpha_{0}+i\approx\alpha_{0}$. This means, if we choose a
relatively large $\alpha_{0}$, we can neglect the imaginary part of
$A_{12}$ ($A_{21}$). This would make sense
because a pulse with form of $\alpha_{0}\dot{\theta}$ would be
more easily to realize than the form of $i\dot{\theta}$ in
experiment. We plot Fig. \ref{fig1} (c) which shows
the result when $\beta=0$ (the other parameters are also
$\{\varphi=0,\ \zeta=3\Omega_{0},\ t_{f}=1/\Omega_{0}$\}). From the
figure, we find the population transfer would be ideally achieved
as long as $\alpha_{0}>2.5$.

\begin{figure}
 \scalebox{0.6}{\includegraphics {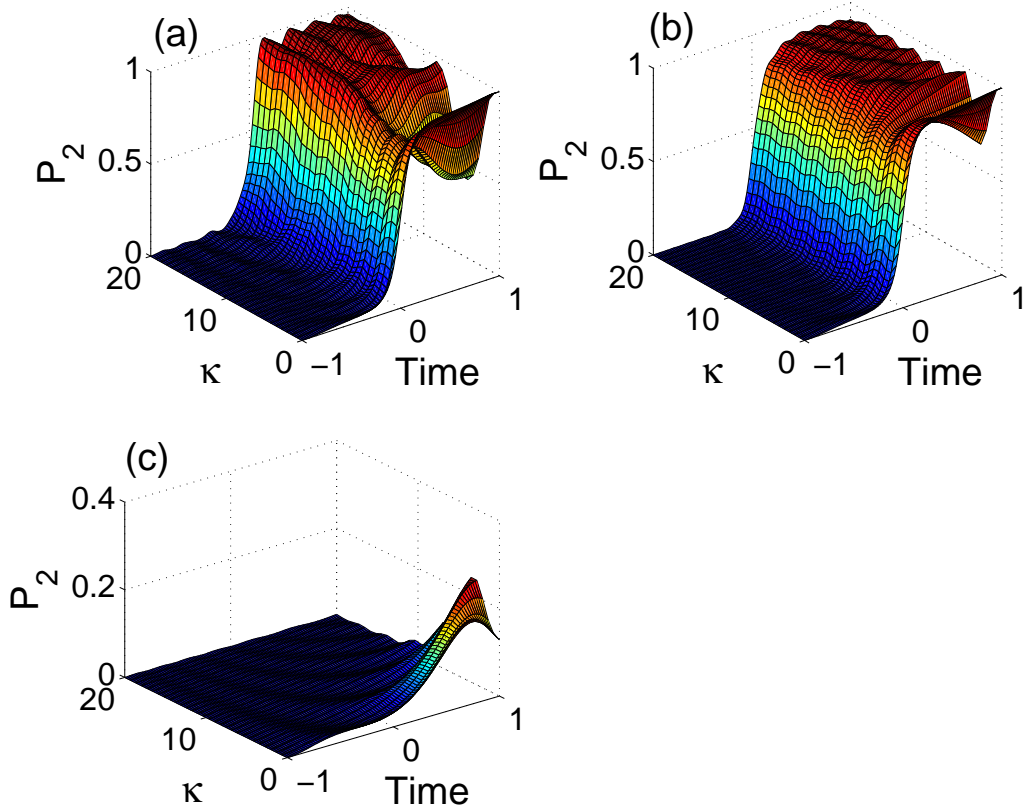}}
 \caption{
    The time-dependent $P_{2}$ versus $\kappa$ (in units $\Omega_{0}$)  when $\{\varphi=\kappa t,\ \zeta=3\Omega_{0},\ t_{f}=1/\Omega_{0}\}$:
    (a) based on the original transitionless tracking algorithm that $\alpha=-(\kappa/2)\sin{2\theta}$;
    (b) based on the present method with parameter $\alpha=0$;
    (c) based on $H_{0}$ without the adding term.
    The evolution time in the figure is in units of $1/\Omega_{0}$.}
 \label{fig2}
\end{figure}

In the following, we will analyze the effectivity of the method when
$\varphi\neq0$. In Fig. \ref{fig2} (a), we give $P_{2}$ versus
$\kappa$ when the initial state is $|1\rangle$ and
$\{\alpha=-\frac{\dot{\varphi}}{2}\sin{2\theta},\ \varphi=\kappa t,\
\zeta=3\Omega_{0}\}$. As shown in the figure, when $t=t_{f}$, while
oscillating, the fidelity of the target state $|2\rangle$ increases
with $\kappa$'s increasing. Which means if the adiabatic phase is
considered, the effectivity of STA may reduce in some situation. For
comparison, in Fig. \ref{fig2} (b), we plot the time-evolution of
state $|2\rangle$ versus $\kappa$ with $\{\alpha=0,\ \varphi=\kappa
t,\ \zeta=3\Omega_{0}\}$. It is obvious that the second set of
parameters behave better in restraining the adverse effect caused by
$\varphi$ than the first set. The oscillation in Figs. \ref{fig2}
(a)  and (b) is caused by the original Hamiltonian $H_{0}$ when
$\Delta$ is large enough as shown in Fig. \ref{fig2} (c). In
addition, it is not hard to find that using the second set of
parameters to construct shortcut can save more energy. According to
eq. (\ref{eq29}), the eigenvalue of $H_{add}$ is
$E^{a}_{\pm}=\pm\hbar\sqrt{(\dot{\varphi}/2-\alpha\cot{2\theta})^2+\alpha^2+\dot{\theta}^{2}}$.
This means the energy cost for constructing shortcuts is the least
when $\alpha=0$.

\begin{figure}
 \scalebox{0.6}{\includegraphics {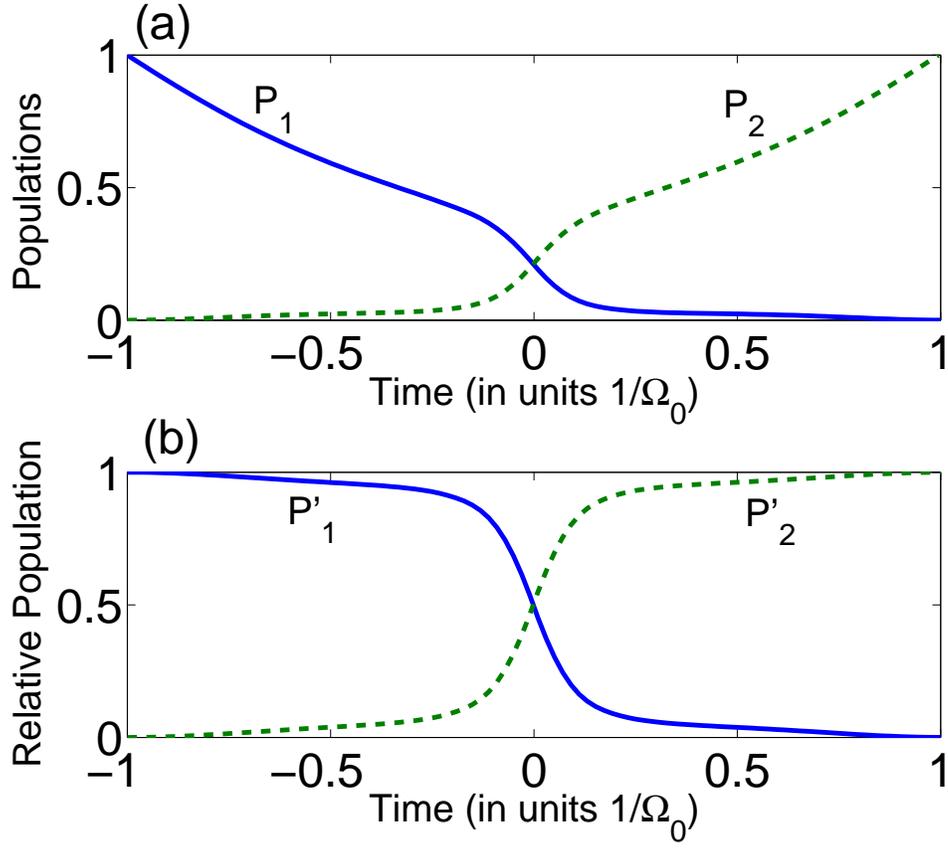}}
 \caption{
    (a) The populations $P_{1}$ and $P_{2}$ versus time when $\gamma=0.5\Omega_{0}$.
    (b) The relative population $P'_{1}$ and $P'_{2}$ versus time when $\gamma=0.5\Omega_{0}$.}
 \label{fig3a}
\end{figure}

\begin{figure}
 \scalebox{0.4}{\includegraphics {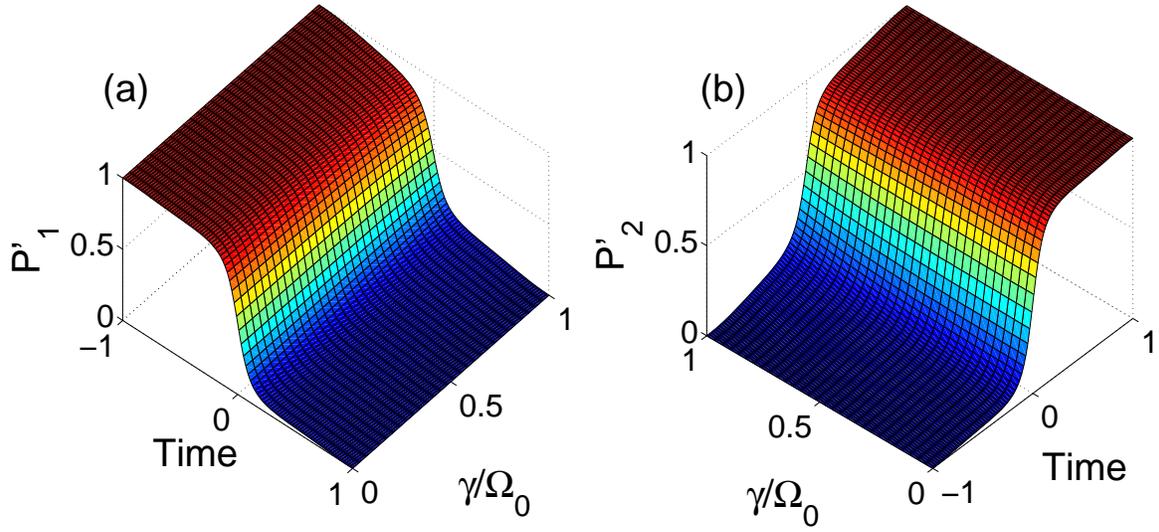}}
 \caption{
    (a) The time-dependent relative population $P'_{1}$ versus $\gamma$.
    (b) The time-dependent relative population $P'_{2}$ versus $\gamma$.
    The evolution time in the figure is in units of $1/\Omega_{0}$.}
 \label{fig3}
\end{figure}

\begin{figure}
 \scalebox{0.6}{\includegraphics {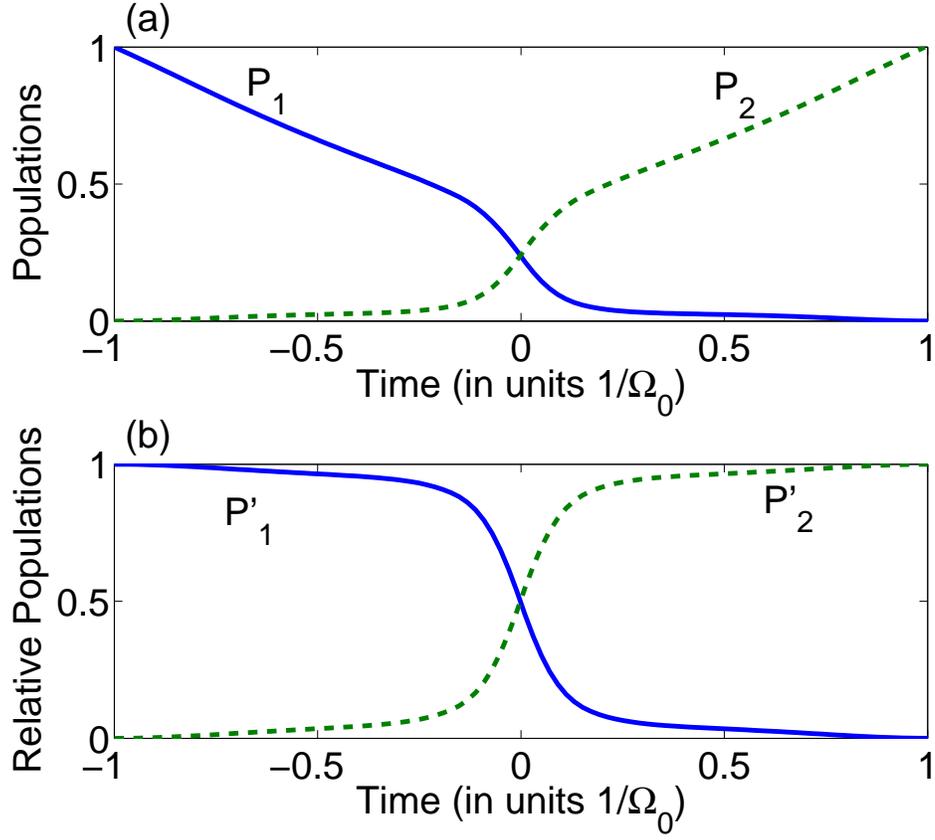}}
 \caption{
    (a) The populations $P_{1}$ and $P_{2}$ versus time when $\gamma=1/(2+t^{2})$.
    (b) The relative population $P'_{1}$ and $P'_{2}$ versus time when $\gamma=1/(2+t^{2})$.}
 \label{fig3b}
\end{figure}

In the following, we will briefly discuss the present method's efficiency in the non-Hermitian system.
Since the system is non-Hermitian, the dynamics of the system's density operator $\rho(t)$ will be given as
$\frac{d}{dt}{\rho(t)}=\frac{1}{i\hbar}[H(t)\rho(t)-\rho(t)H^{\dag}(t)]$, where $H(t)=H_{0}(t)+H_{add}(t)$.
First of all, we assume the population for a state $|j\rangle$ is still given as $P_{j}=|\langle j|\rho(t)|j\rangle|$,
and display the populations $P_{1}$ and $P_{2}$ versus time in Fig. \ref{fig3a} with parameters
$\{\alpha=0,\ \varphi=0,\ \zeta=3\Omega_{0}, \gamma=0.5\Omega_{0}\}$.
%For simplicity, we assume $\gamma$ is time-independent.
%Figs. \ref{fig3} (a) and (b) show the relationship between the time-dependent populations of $|1\rangle$ and $|2\rangle$ verus $\gamma$
%when $\{\alpha=0,\ \varphi=0,\ \zeta=3\Omega_{0}\}$.
It should be noted here, since the Hamiltonian is non-Hermitian,
if the population for a state is still given by $P_{j}=|\langle j|\rho(t)|j\rangle|$, the
norm of the state vector given by $P_{1}+P_{2}$ will not
be conserved during the evolution. This property can be seen
in Fig. \ref{fig3a}, where the norm is not conserved during the interaction.
To avoid some problems caused by $P_{1}+P_{2}\neq 1$,
some definitions of population in non-Hermitian system have been proposed \cite{Jpa45415201,Pra89033403}.
However, since we only concern about the realizable possibility of the fast population inversion in the non-hermitian system,
for simplicity, we define relative populations $P_{j}'=P_{j}/(P_{j}+P_{k})$ ($j\neq k$) to help to analyze, and
if no otherwise specified, $\alpha=0$, $\varphi=0$, and $\zeta=3\Omega_{0}$ will be used throughout the discussion in this part.
%We plot the relative populations $P'_{1,2}$ versus time in Fig. \ref{fig3a} (b) also with parameters
%$\{\alpha=0,\ \varphi=0,\ \zeta=3\Omega_{0}, \gamma=0.5\Omega_{0}\}$ for comparison.
In Fig. \ref{fig3} we display the time-dependent relative
populations for states $|1\rangle$ [Fig. \ref{fig3} (a)] and
$|2\rangle$ [Fig. \ref{fig3} (b)] versus $\gamma$, where $\gamma$ is
assumed time-independent. As we can see, the fast population
inversion still could be achieved even with a relative large
$\gamma$, i.e., $\gamma=\Omega_{0}$. As it is known, in
general, $\gamma$ could also depend on time,
$\gamma=\gamma(t)$, as an effective decay rate controlled by further
interactions [see, e.g., ref. \cite{Jpb41175501}]. According to
the form of $\gamma$ in ref. \cite{Jpb41175501}, we plot Fig.
\ref{fig3b} to show that the present method can also work very well
in the case of $\gamma$ is time-dependent, which shows the
populations versus time with the parameters mentioned above. Fig. \ref{fig3b} (b) shows the relative populations versus time,
and $\gamma$ is chosen as $\gamma=\frac{1}{2+t^{2}}$ for simplicity
in plotting the figures. Moreover, if $\gamma$ is controllable, or
if $\gamma$ could satisfy some kind of function, for example,
$\gamma=\pm\dot{\theta}/\sin{2\theta}$, the scheme can make
the population transfer fast without increasing the coupling
\cite{Pra86033813} because when
$\gamma=\pm\dot{\theta}/\sin{2\theta}$, the corresponding $\beta=0$.
Such assumption can be physically realized, for instance, in two
coupled optical waveguides with longitudinally varying gain and loss
regions \cite{Pra86033813}. In fact,
$\gamma=\pm\dot{\theta}/\sin{2\theta}$ is just the result of ref.
\cite{Pra8705250289063412} which has been analyzed and
discussed in very detail.

\begin{figure}
 \scalebox{0.6}{\includegraphics {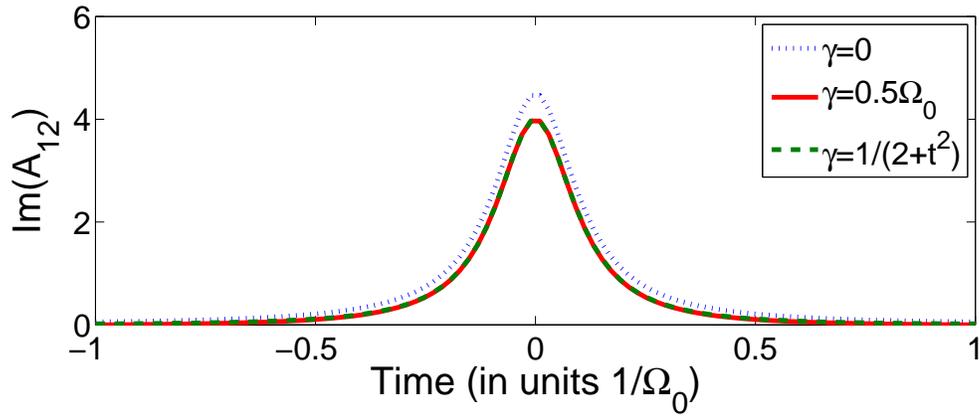}}
 \caption{
    The shapes of the imaginary part of the adding Hamiltonian's pulses when $\varphi=0$. Blue dotted curve when $\gamma=0$; Red solid curve when $\gamma=0.5\Omega_{0}$; Green dashed curve when  $\gamma=1/(2+t^{2})$.}
 \label{fig4}
\end{figure}

From the analysis above, we find the real part of pulse Re$A_{12}$ could be arbitrary time-dependent function, which
means the real part is obviously realizable.
So, to make sure the pulses we used in the schemes are realizable, we need to confirm that whether the imaginary part of the pulse is realizable or not.
Fig. \ref{fig4} shows Im$A_{12}$ versus time with different parameters when $\varphi=0$.
Shown in the figure, the shapes are all similar to Gaussian curves, which means
the pulses are not hard to be realized in practice. In other words, the schemes proposed in the paper
are feasible in practice.

In conclusion, we have proposed a different and flexible way to
design the adding Hamiltonian for the original Hamiltonian to
construct shortcuts to adiabaticity (STA). The method maybe
promising to avoid the trouble (the speed-up protocols' Hamiltonian
may involve the terms which are difficult to be realized in
practice) because of the multiple-choices of the adding Hamiltonian.
We have applied this method to the Landau-Zener model as an
application example, and the results show the method works very well
in two-level systems (in both Hermitian and non-Hermitian). In
Hermitian system, we find a relatively suitable $\alpha$ (the
real part of the off-diagonal terms in the adding Hamiltonian), we
can even speed up the adiabatic process without the imaginary part
of the off-diagonal terms in the adding Hamiltonian. That is meaningful
because amending the Rabi frequency $\Omega$ by real
correction will be much more easily than by imaginary
correction. In non-Hermitian system, different from ref.
\cite{Pra8705250289063412} where $\gamma$ (gain or loss of
population) nullifies the counterdiabatic coupling to speed up the
adiabatic evolution all alone, in this paper, $\gamma$ cooperates
with $\beta$ (the correction of the imaginary part of Rabi
frequency) to achieve the goals. As is known, the decay $\gamma$ is
usually decided by the system and is uncontrollable, so a
speed-up protocol with a fixed form of $\gamma$ will be hard to
realize and generalize. However, in our present method, the
correction of Rabi frequency $\beta$ cooperates with $\gamma$ to
construct shortcuts, hence, as long as the corresponding $\beta$ is
realizable in practice, the shortcuts could be constructed
with arbitrary $\gamma$. Another highlight of this method is that
the phase change at any time could be obviously calculated which may
have application prospect in quantum phase gates.
%The feature is that the method might be hard to deal with multi-level and multi-particle systems because (1)
%the nonadiabatic coupling will be hard to calculated when the system is complicated; (2) there will be too many unknowns in the adding Hamiltonian that
%the calculating amount will be quite large.

\section*{ACKNOWLEDGEMENT}

  This work was supported by the National Natural Science Foundation of China under Grants No.
11575045 and No. 11374054, and the Major State Basic Research
Development Program of China under Grant No. 2012CB921601.

\newpage

\end{document}